\documentclass[11pt,a4paper,fleqn]{article}

\usepackage{amsmath,amssymb,amsthm,enumerate,cite}

\newcommand{\p}{\partial}
\newcommand{\const}{\mathop{\rm const}\nolimits}

\newcommand{\Equiv}{\mathop{\sim}}

\newtheorem{theorem}{Theorem}
\newtheorem{lemma}{Lemma}
\newtheorem{corollary}{Corollary}
\newtheorem{proposition}{Proposition}
{\theoremstyle{definition} \newtheorem{definition}{Definition}

\begin{document}

\par\noindent {\LARGE\bf Potential Nonclassical Symmetries and\\  Solutions of Fast Diffusion Equation
\par}
{\vspace{4mm}\par\noindent {\it
Roman~O.~Popovych~$^\dag$, Olena~O.~Vaneeva~$^\ddag$ and Nataliya~M.~Ivanova~$^\S$
} \par\vspace{2mm}\par}
{\vspace{2mm}\par\noindent {\it
${}^\dag{}^\ddag{}^\S$Institute of Mathematics of National Academy of Sciences of Ukraine, \\
$\phantom{{}^\dag{}^\ddag{}^\S}$3 Tereshchenkivska Str., Kyiv-4, 01601 Ukraine\\
$\phantom{{}^\dag{}^\ddag{}}{}^\dag$~Fakult\"at f\"ur Mathematik, Universit\"at Wien, Nordbergstra{\ss}e 15, A-1090 Wien, Austria\\
$\phantom{{}^\dag{}^\ddag{}}{}^\S$Department of Mathematics, UBC, Vancouver, BC, V6T 1Z2, Canada
}}
\par\vspace{2mm}\par
{\noindent$\phantom{{}^\dag{}^\ddag{}}$ E-mail: {\it
${}^\dag$rop@imath.kiev.ua, $^\ddag$vaneeva@imath.kiev.ua, $^\S$ivanova@imath.kiev.ua
} \par}

{\vspace{5mm}\par\noindent\hspace*{5mm}\parbox{150mm}{\small
The fast diffusion equation $u_t=(u^{-1}u_x)_x$ is investigated from the symmetry point of view
in development of the paper by Gandarias [Phys.~Lett.~A 286 (2001) 153--160].
After studying equivalence of nonclassical symmetries with respect to a
transformation group, we completely classify the nonclassical symmetries of the corresponding potential equation.
As a result, new wide classes of potential nonclassical symmetries of the fast diffusion equation are obtained.
The set of known exact non-Lie solutions are supplemented with the similar ones.
It is shown that all known non-Lie solutions of the fast diffusion equation are exhausted by ones
which can be constructed in a regular way with the above potential nonclassical symmetries.
Connection between classes of nonclassical and potential nonclassical symmetries of
the fast diffusion equation is found.
}\par\vspace{5mm}}

\section{Introduction}
Investigation of nonlinear heat (or diffusion if $u$
represents mass concentration) equations by means of symmetry methods was started as early as
in 1959 with Ovsiannikov's work~\cite{Ovsiannikov1959}
where the author performed the group classification of the class of equations of the form
\begin{equation}\label{EqNonlinDiff}
u_t=(f(u)u_x)_x.
\end{equation}

Nonclassical symmetries of equations from class~\eqref{EqNonlinDiff} were investigated
in~\cite{Amerov1990,Fushchych&Serov&Amerov1982,Fushchych&Serov&Tulupova1993,Gandarias2001}.
In particular, the authors of~\cite{Fushchych&Serov&Tulupova1993} obtained the determining equations
for the coefficients of conditional symmetry operators for the wider class of nonlinear reaction--diffusion equations
of the form~$u_t=(f(u)u_x)_x+g(u)$ and constructed a number of their exact solutions.
Review of results on symmetries, exact solutions and conservation laws of such equations is given e.g.
in~\cite{Ibragimov1994V1}.

The diffusion processes described by~\eqref{EqNonlinDiff} are known to arise in different fields
of physics such as plasma physics, kinetic theory of gases, solid state and transport in porous medium.
In many metals and ceramic materials the diffusion coefficient $f(u)$ can, over a wide range of temperatures,
be approximated as $u^{-\alpha}$, where $0<\alpha<2$~\cite{Rosenau1995}.
So, one of the mathematical model of the diffusion processes is
\begin{equation}\label{diffusionmodel}
u_t=(u^{-\alpha}u_x)_x.
\end{equation}
Equations~\eqref{diffusionmodel} are called {\it fast diffusion
equations} in the case $0<\alpha<2$ since these values of~$\alpha$
correspond to a much faster spread of mass than in the linear case
($\alpha=0$).

In this Letter we restrict ourselves with the special case
$\alpha=1$, i.e. with the equation
\begin{equation} \label{FastDifEq}
u_t=(u^{-1}u_x)_x.
\end{equation}
It emerges in plasma physics as a model of the cross-field convective diffusion of
plasma including mirror effects and in the central limit approximation to Calerman's model of the Boltzmann equation.
Equation~\eqref{FastDifEq} governs the expansion of a thermalized
electron cloud described by isothermal Maxwell distribution. It is also the one-dimensional Ricci flow equation.
(See~\cite{Bakas2004,Berryman&Holland1982,Rosenau1995} and references therein.)

Equation~\eqref{FastDifEq} has a number of remarkable mathematical properties
which distinguish it from class~\eqref{diffusionmodel}.
Thus, \eqref{FastDifEq} can be rewritten in the form $u_t=(\ln u)_{xx}$ whereas for the other values of~$\alpha$
the function under $\p_{xx}$ is a power one. It admits a discrete potential invariance transformation.
For this equation wide classes of exact solutions were constructed in a closed form
while reduction of~\eqref{diffusionmodel} in the general case results in ordinary differential equations which
usually cannot be integrated explicitly. Its potential form admits two kinds of variable separation.

The fact that~\eqref{FastDifEq} is written in a conserved form allows us,
following Bluman et al.~\cite{Bluman&Kumei1989,Bluman&Reid&Kumei1988,Bluman&Yan2005}, to consider the corresponding
 potential  system
\begin{equation} \label{PotSysFastDif}
v_x=u, \quad v_t=u^{-1}u_x
\end{equation}
and to find potential symmetries of equation~\eqref{FastDifEq}.
Namely, any local symmetry of system~\eqref{PotSysFastDif} induces a symmetry of the initial equation~\eqref{FastDifEq}.
If transformations of some of the ``non-potential'' variables $t$, $x$ and $u$ explicitly depend on the potential $v$,
this symmetry is a nonlocal (potential) symmetry of equation~\eqref{FastDifEq}.

It follows from \eqref{PotSysFastDif} that the potential~$v$ satisfies the nonlinear filtration equation
\begin{equation} \label{PotFastDifEq}
v_t=v_x{}^{-1}v_{xx}
\end{equation}
with the special value $v_x{}^{-1}$ of the filtration coefficient.
We will also call equation~\eqref{PotFastDifEq} the {\it potential
fast diffusion equation}. Akhatov, Gazizov and
Ibragimov carried out group classification of the nonlinear
filtration equations of the general form
\begin{equation}\label{EqNonlinFiltration}
v_t=f(v_x)v_{xx}
\end{equation}
and investigated their contact and quasi-local
symmetries~\cite{Akhatov&Gazizov&Ibragimov1987,Akhatov&Gazizov&Ibragimov1989,Ibragimov1994V1}.

Lie symmetries of~\eqref{FastDifEq} are well known (see
Section~2). All its exact solutions constructed in closed form by
reduction with Lie symmetries are listed e.g.
in~\cite{Polyanin&Zaitsev2004}.

Some non-Lie exact solutions of~\eqref{FastDifEq} were obtained in~\cite{Rosenau1995,Qu1999,Gandarias2001}.
Thus, Rosenau~\cite{Rosenau1995} found that equation~\eqref{PotFastDifEq} admits,
in addition to the usual variable separation $v=T(t)X(x)$, the additive one $v=Y(x+\lambda t)+Z(x-\lambda t)$
which is a potential additive variable separation for equation~\eqref{FastDifEq}.
To construct nonclassical solutions of~\eqref{FastDifEq},
Qu~\cite{Qu1999} made use of generalized conditional symmetry method, looking for the conditional symmetry
operators in the special form $Q=(u_{xx}+H(u){u_x}^2+F(u)u_x+G(u))\p_u$.
Gandarias~\cite{Gandarias2001} investigated some families of usual and potential nonclassical symmetries
of~\eqref{diffusionmodel}. In particular, using an ansatz for the coefficient~$\eta$, she found
nontrivial reduction operators in the so-called ``no-go''
case~\cite{Fushchych&Shtelen&Serov&Popovych1992,Zhdanov&Lahno1998,Popovych1998}
when the coefficient of~$\p_t$ vanishes,
i.e. operators can be reduced to the form~$Q=\p_x+\eta(t,x,u)\p_u$.

In the recent paper~\cite{Bluman&Yan2005} a preliminary analysis
of nonclassical symmetries of equations from
class~\eqref{EqNonlinFiltration} was performed. A more detailed
consideration was carried out for the case
\mbox{$f=(v_x^2+v_x)^{-1}$}, and only some examples of reduction
operators and corresponding exact solutions were constructed. Let
us note that equation~\eqref{EqNonlinFiltration} with
$f=(v_x^2+v_x)^{-1}$ is reduced by the point transformation
$\tilde t=t$, $\tilde x=x+v$, $\tilde v=v$ to
equation~\eqref{PotFastDifEq} which corresponds to the value
$\tilde f=\tilde v_{\tilde x}{}^{-1}$ and is simpler and more
convenient for investigation. All results on symmetries and exact
solutions of the equation from~\cite{Bluman&Yan2005} can be
derived from the analogous results for
equation~\eqref{PotFastDifEq}.

In this Letter the fast diffusion equation~\eqref{FastDifEq} is
investigated from the symmetry point of view. The nonclassical
symmetries of the corresponding potential
equation~\eqref{PotFastDifEq} are completely classified with
respect to its Lie symmetry group. As a result, new wide classes
of potential nonclassical symmetries of equation~\eqref{FastDifEq}
are found. Some classes of potential nonclassical symmetries prove
to be connected with usual nonclassical ones on the solution set
of potential system~\eqref{PotSysFastDif}. The set of exact
non-Lie solutions constructed
in~\cite{Rosenau1995,Qu1999,Gandarias2001} is supplemented with
the similar ones. It~is shown that all known non-Lie solutions of
the fast diffusion equation are exhausted by ones which can be
constructed with the above potential nonclassical symmetries.

Our Letter is organized as follows. First of all (Section~\ref{SectionOnLiesymmetriesOfFastDifEq})
we adduce results on Lie and potential symmetries of~\eqref{FastDifEq}, including discrete ones.
It is important since classical symmetries really are partial cases of nonclassical symmetries and
below we solve the problem on finding only pure nonclassical symmetries which are not equivalent to classical ones.
Moreover, our approach is based on application of the notion of equivalence of nonclassical symmetries
with respect to a transformation group, which is developed and investigated
in Section~\ref{SectionOnEquivOfRedOpsWrtTransGroup}.
Usage of equivalence with respect to the complete Lie invariance group including the discrete symmetries
plays a significant role in simplification of proof, testing and improving presentation of the main result
(Theorem~\ref{TheoremOnRedOpsOfPotFastDifEq}, Section~\ref{SectionOnRedOpsOfPotFastDifEq}).
In spite of the techniques applied in~\cite{Gandarias2001} and similarly to~\cite{Bluman&Yan2005},
we use the single potential equation~\eqref{PotFastDifEq} instead of potential system~\eqref{PotSysFastDif},
to produce potential nonclassical symmetries of~\eqref{FastDifEq}.
After the ``no-go'' case of the zero coefficient of $\p_t$ is discussed,
all the reduction operators having the nonvanishing coefficient of $\p_t$ are classified.
Connection between partial classes of usual and potential reduction operators of~\eqref{EqNonlinDiff}
is studied in Section~\ref{SectionOnConnectionBetweenUsualAndPotRedOpsOfFastDifEq}.
In Section~\ref{SectionOnSolutionOfFastDifEq} the known Lie solutions of~\eqref{FastDifEq} and~\eqref{PotFastDifEq}
are collected. A list of non-Lie solutions is supplemented with the similar ones.
Connections between exact solutions and different ways of their construction are discussed shortly.
In conclusion some recent results on nonclassical symmetries of equations~\eqref{EqNonlinFiltration} are announced.

\section{Lie and potential symmetries of fast diffusion equation}\label{SectionOnLiesymmetriesOfFastDifEq}

The Lie invariance algebra
\[
A_1=\langle \p_t,\; \p_x,\; t\p_t+u\p_u,\; x\p_x-2u\p_u\rangle
\]
of equation~\eqref{FastDifEq} was found in~\cite{Ovsiannikov1959}.
The complete Lie invariance group~$G_1$ of~\eqref{FastDifEq} is generated by both
continuous one-parameter transformation groups with infinitesimal operators from $A_1$ and
two involution transformations of alternating sign in the sets $\{t,u\}$ and $\{x\}$.
Action of any element from~$G_1$ on the function~$u$ is given by the formula
\[
\tilde u(t,x)=\varepsilon_3^{-1}\varepsilon_4^2\,u(\varepsilon_3t+\varepsilon_1,\varepsilon_4x+\varepsilon_2),
\]
where $\varepsilon_1$, \dots, $\varepsilon_4$ are arbitrary constants,
$\varepsilon_3\varepsilon_4\ne0$~\cite{Ovsiannikov1959}.

\looseness=-1
The Lie symmetry properties of~\eqref{FastDifEq} are common for diffusion equations.
Uncommonness of equation~\eqref{FastDifEq} from the symmetry point of view becomes apparent after introducing
the potential~$v$ and considering potential system~\eqref{PotSysFastDif} or potential equation~\eqref{PotFastDifEq}.
Point and nonclassical symmetries of~\eqref{PotSysFastDif} or~\eqref{PotFastDifEq} are called
\emph{potential} and \emph{potential nonclassical} symmetries of~\eqref{FastDifEq} correspondingly.

The Lie invariance algebra
\[
A_2=\langle\p_t,\,\p_x,\,\p_v,\,t\p_t+v\p_v,\,x\p_x-v\p_v\rangle
\]
of equation~\eqref{PotFastDifEq} and the corresponding connected
Lie symmetry group are quite ordinary for nonlinear filtration
equations. However, equation~\eqref{PotFastDifEq} is distinguished
for its discrete symmetries since it possesses, besides two usual
sign changes in the variable sets $\{t,v\}$ and $\{x,v\}$, the
hodograph transformation  $\tilde t=t$, $\tilde x=v$, $\tilde
v=x$. These three involutive transformations together with the
continuous one-parameter transformation groups having
infinitesimal operators from $A_2$ generate the complete Lie
invariance group~$G_2$ of~\eqref{PotFastDifEq}. Therefore, $G_2$
consists of the transformations
\[\begin{array}{l}
\tilde t=\varepsilon_3t+\varepsilon_1,\quad
\tilde x=\varepsilon_4x+\varepsilon_2,\quad
\tilde v=\varepsilon_3{\varepsilon_4}^{-1}v\quad\mbox{and}\\
\tilde t=\varepsilon_3t+\varepsilon_1,\quad \tilde
x=\varepsilon_3{\varepsilon_4}^{-1}v,\quad \tilde
v=\varepsilon_4x+\varepsilon_2,
\end{array}\]
where $\varepsilon_1$, \dots, $\varepsilon_4$ are arbitrary
constants, $\varepsilon_3\varepsilon_4\ne0$.

A similar result is true for system~\eqref{PotSysFastDif}.
Namely, it is invariant with respect to the following transformation
\begin{equation}\label{pothodograph}
\tilde t=t,\quad
\tilde x=v,\quad
\tilde u=u^{-1},\quad
\tilde v=x
\end{equation}
which is additional to the usual Lie symmetry group~$G_1$ of equation~\eqref{FastDifEq}
and is called the potential hodograph transformation of this equation.

It can be proved~\cite{Popovych&Ivanova2005PETs} that
the set of Lie invariant solutions of equation~\eqref{FastDifEq} is closed under transformation~\eqref{pothodograph}.

\section{Equivalence of reduction operators\\ with respect to transformation groups}
\label{SectionOnEquivOfRedOpsWrtTransGroup}

The notion of nonclassical symmetry was introduced
 in 1969~\cite{Bluman&Cole1969}. A precise and rigorous definition
was suggested later (see
e.g.~\cite{Fushchych&Tsyfra1987,Zhdanov&Tsyfra&Popovych1999}).

Consider an $r$th order differential equation~$\mathcal{L}$ of the form~$L(t,x,u_{(r)})=0$
for the unknown function $u$ of two independent variables~$t$ and~$x$.
Here $u_{(r)}$ denotes the set of all the derivatives of the function $u$ with respect to $t$ and~$x$
of order not greater than~$r$, including $u$ as the derivative of order zero.
Within the local approach the equation~$\mathcal{L}$ is treated as an algebraic equation
in the jet space $J^{(r)}$ of order $r$ and is identified with the manifold of its solutions in~$J^{(r)}$:
\begin{gather*}
{\cal L}=\{ (t,x,u_{(r)}) \in J^{(r)}\, |\, L(t,x,u_{(r)})=0\}.
\end{gather*}

The set of (first-order) differential operators of the general form
\[
Q=\tau(t,x,u)\p_t+\xi(t,x,u)\p_x+\eta(t,x,u)\p_u, \quad (\tau,\xi)\not=(0,0),
\]
will be denoted by ${\cal Q}$.
Here and below $\p_t=\p/\p t$, $\p_x=\p/\p x$ and $\p_u=\p/\p u$.
Subscripts of functions denote differentiation with respect to the corresponding variables.

Two differential operators $\widetilde Q$ and $Q$ are called \emph{equivalent} if they differ by a multiplier
being a non-vanishing function of~$t$, $x$ and~$u$:
$\widetilde Q=\lambda Q$, where $\lambda=\lambda(t,x,u)$, $\lambda\not=0$.
The equivalence of operators will be denoted by $\widetilde Q\sim Q$.
Denote also the result of factorization of~${\cal Q}$ with respect to this equivalence relation by~${\cal Q}_{\rm f}$.
Elements of~${\cal Q}_{\rm f}$ will be identified with their representatives in~${\cal Q}$.

The first-order differential function~$Q[u]:=\eta(t,x,u)-\tau(t,x,u)u_t-\xi(t,x,u)u_x$
is called the {\it characteristic} of the operator~$Q$.
The characteristic PDE $Q[u]=0$ (called also the \emph{invariant surface condition}) has
two functionally independent integrals $\zeta(t,x,u)$ and $\omega(t,x,u)$.
Therefore, the general solution of this equation can be implicitly presented in the form
$F(\zeta,\omega)=0$, where~$F$ is an arbitrary function of its arguments.

The characteristic equations of equivalent operators have the same set of solutions.
And vice versa, any family of two functionally independent functions of~$t$, $x$ and $u$
is a complete set of integrals of the characteristic equation of a differential operator.
Therefore, there exists a one-to-one correspondence between ${\cal Q}_{\rm f}$
and the set of families of two functionally independent functions of~$t$, $x$ and $u$,
which is factorized with respect to the corresponding equivalence.
(Two families of the same number of functionally independent functions of the same arguments are considered equivalent
if any function from one of the families is functionally dependent on functions from the other family.)

Since $(\tau,\xi)\not=(0,0)$ we can assume without loss of generality that
$\zeta_u\not=0$ and $F_\zeta\not=0$ and resolve the equation~$F=0$ with respect to~$\zeta$: $\zeta=\varphi(\omega)$.
This implicit representation of the function~$u$ is called an \emph{ansatz} corresponding to the operator~$Q$.

Denote the manifold defined by the set of all the differential
consequences of the characteristic equation~$Q[u]=0$ in $J^{(r)}$
by ${\cal Q}^{(r)}$, i.e.
\begin{gather*}
{\cal Q}^{(r)}=\{ (t,x,u_{(r)}) \in J^{(r)}\, |\,
D_t^{\alpha}D_x^{\beta}Q[u]=0, \
\alpha,\beta\in\mathbb{N}\cup\{0\},\ \alpha+\beta<r \},
\end{gather*}
where $D_t=\p_t+u_{\alpha+1,\beta}\p_{u_{\alpha\beta}}$ and
$D_x=\p_x+u_{\alpha,\beta+1}\p_{u_{\alpha\beta}}$ are the
operators of total differentiation with respect to the
variables~$t$ and~$x$, the variable $u_{\alpha\beta}$ of the jet
space $J^{(r)}$ corresponds to the derivative
$\p^{\alpha+\beta}u/\p t^{\alpha}\p x^{\beta}$.

\begin{definition}\label{DefinitionOfCondSym}
The differential equation~$\mathcal{L}$ is called
\emph{conditionally invariant} with respect to the operator $Q$ if
the relation
$Q_{(r)}L(t,x,u_{(r)})\bigl|_{\mathcal{L}\cap\mathcal{Q}^{(r)}}=0$
holds, which is called the \emph{conditional invariance
criterion}. Then $Q$ is called an operator of \emph{conditional
symmetry} (or $Q$-conditional symmetry, nonclassical symmetry, etc.)
of the equation~$\mathcal{L}$.
\end{definition}

In Definition~\ref{DefinitionOfCondSym} the symbol $Q_{(r)}$ stands for the standard $r$th prolongation
of the operator~$Q$ \cite{Olver1986,Ovsiannikov1982}:
$Q_{(r)}=Q+\sum_{0<\alpha+\beta{}\leqslant  r} \eta^{\alpha\beta}\p_{u_{\alpha\beta}}$,
where
$\eta^{\alpha\beta}=D_t^{\alpha}D_x^{\beta}Q[u]+\tau u_{\alpha+1,\beta}+\xi u_{\alpha,\beta+1}$.

The equation~$\mathcal{L}$ is conditionally invariant with respect to the operator~$Q$
iff the ansatz constructed with this operator reduces~$\mathcal{L}$
to an ordinary differential equation~\cite{Zhdanov&Tsyfra&Popovych1999}.
So, we will also call operators of conditional symmetry by {\it reduction operators} of~$\mathcal{L}$.

\begin{lemma}[\cite{Fushchych&Zhdanov1992,Zhdanov&Tsyfra&Popovych1999}]\label{LemmaOnEquivFamiliesOfOperators}
If the equation~$\mathcal{L}$ is conditionally invariant with respect to the operator~$Q$,
then it is conditionally invariant with respect to any operator which is equivalent to~$Q$.
\end{lemma}

The set of reduction operators of the equation~$\mathcal{L}$ is a
subset of ${\cal Q}$ and so will be denoted by ${\cal Q}({\cal
L})$. In view of Lemma~\ref{LemmaOnEquivFamiliesOfOperators},
$Q\in {\cal Q}({\cal L})$ and $\widetilde Q\sim Q$ imply
$\widetilde Q\in {\cal Q}({\cal L})$, i.e. ${\cal Q}({\cal L})$ is
closed under the equivalence relation on ${\cal Q}$. Therefore,
factorization of ${\cal Q}$ with respect to this equivalence
relation can be naturally restricted on~${\cal Q}({\cal L})$ that
results in the subset~${\cal Q}_{\rm f}({\cal L})$ of ${\cal
Q}_{\rm f}$. As in the whole set~${\cal Q}_{\rm f}$, we identify
elements of~${\cal Q}_{\rm f}({\cal L})$ with their
representatives in~${\cal Q}({\cal L})$. In this approach the
problem of complete description of reduction operators for the
equation~$\mathcal{L}$ is nothing but the problem of
finding~${\cal Q}_{\rm f}({\cal L})$.

We can essentially simplify and order classification of reduction operators, additionally taking into account
Lie symmetry transformations of an equation or equivalence transformations of a whole class of equations.

\begin{lemma}
Any point transformation of $t$, $x$ and $u$ induces a one-to-one mapping of~${\cal Q}$ into itself.
Namely, the transformation~$g$: $\tilde t=T(t,x,u)$, $\tilde x=X(t,x,u)$, $\tilde u=U(t,x,u)$ generates
the mapping~\mbox{$g_*\colon {\cal Q}\to{\cal Q}$} such that
the operator~$Q$ is mapped to the operator
$g_*Q=\tilde\tau\p_{\tilde t}+\tilde\xi\p_{\tilde x}+\tilde\eta\p_{\tilde u}$, where
$\tilde\tau(\tilde t,\tilde x,\tilde u)=QT(t,x,u)$,
$\tilde\xi(\tilde t,\tilde x,\tilde u)=QX(t,x,u)$,
$\tilde\eta(\tilde t,\tilde x,\tilde u)=QU(t,x,u)$.
If~$Q'\sim Q$ then  $g_* Q'\sim g_* Q$.
Therefore, the corresponding factorized mapping~$g_{\rm f} \colon{\cal Q}_{\rm f}\to{\cal Q}_{\rm f}$ also
is well-defined and one-to-one.

\end{lemma}

\begin{definition}[\cite{Popovych&Korneva1998,Popovych2000}]\label{DefinitionOfEquivInvFamiliesWrtGroup}
The differential operators $Q$ and $\widetilde Q$ are called equivalent
with respect to a group $G$ of point transformations if there exists $g\in G$
for which the operators $Q$ and $g_*\widetilde Q$ are equivalent.
{\it Notation:} $Q\sim \widetilde Q \bmod G.$
\end{definition}

\begin{lemma}\label{LemmaOnInducedMapping}
Given any point transformation $g$ of the equation~$\mathcal{L}$ to an equation~$\tilde{\mathcal{L}}$,
$g_*$ maps~${\cal Q}({\cal L})$ to~${\cal Q}(\tilde{\cal L})$ in a one-to-one manner.
The same statement is true for the factorized mapping $g_{\rm f}$ from ${\cal Q}_{\rm f}({\cal L})$
to~${\cal Q}_{\rm f}(\tilde{\cal L})$.
\end{lemma}

\begin{corollary}\label{CorollaryOnEquivReductionOperatorWrtSymGroup}
Let $G$ be a Lie symmetry group of the equation~$\mathcal{L}.$ Then the equivalence of operators
with respect to the group $G$ generates equivalence relations in~${\cal Q}({\cal L})$
and in~${\cal Q}_{\rm f}({\cal L})$.
\end{corollary}

Consider the class~$\mathcal{L}|_{\cal S}$ of equations~$\mathcal{L}_\theta$: $L(t,x,u_{(r)},\theta(t,x,u_{(r)}))=0$
parameterized with the parameter-functions~$\theta=\theta(t,x,u_{(r)}).$
Here $L$ is a fixed function of $t$, $x$, $u_{(r)}$ and $\theta.$
$\theta$~denotes the tuple of arbitrary (parametric) functions
$\theta(t,x,u_{(r)})=(\theta^1(t,x,u_{(r)}),\ldots,\theta^k(t,x,u_{(r)}))$
running the set~${\cal S}$ of solutions of the system~$S(t,x,u_{(r)},\theta_{(q)}(t,x,u_{(r)}))=0$.
This system consists of differential equations on $\theta$,
where $t$, $x$ and $u_{(r)}$ play the role of independent variables
and $\theta_{(q)}$ stands for the set of all the partial derivatives of $\theta$ of order not greater than $q$.
In what follows we call the functions $\theta$ arbitrary elements.
Denote the point transformations group preserving the
form of the equations from~$\mathcal{L}|_{\cal S}$ by $G^{\Equiv}$.

Let $P$ denote the set of the pairs each of which consists of an
equation $\mathcal{L}_\theta$ from~$\mathcal{L}|_{\cal S}$ and an
operator~$Q$ from~${\cal Q}({\cal L}_\theta)$. In view of
Lemma~\ref{LemmaOnInducedMapping}, action of transformations from
the equivalence group~$G^{\Equiv}$ on $\mathcal{L}|_{\cal S}$ and
$\{{\cal Q}(\mathcal{L}_{\theta})\,|\,\theta\in{\cal S}\}$
together with the pure equivalence relation of differential
operators naturally generates an equivalence relation on~$P$.

\begin{definition}\label{DefinitionOfEquivOfRedOperatorsWrtEquivGroup}
Let $\theta,\theta'\in{\cal S}$,
$Q\in{\cal Q}(\mathcal{L}_\theta)$, $Q'\in{\cal Q}(\mathcal{L}_{\theta'})$.
The pairs~$(\mathcal{L}_\theta,Q)$ and~$(\mathcal{L}_{\theta'},Q')$
are called {\em $G^{\Equiv}$-equivalent} if there exists $g\in G^{\Equiv}$
which transforms the equation~$\mathcal{L}_\theta$ to the equation~$\mathcal{L}_{\theta'}$, and
$Q'\sim g_*Q$.
\end{definition}

Classification of reduction operators with respect to~$G^{\Equiv}$ will be understood as
classification in~$P$ with respect to the above equivalence relation.
This problem can be investigated in the way that is similar to usual group classification in classes
of differential equations.
Namely, we construct firstly the reduction operators that are defined for all values of the arbitrary elements.
Then we classify, with respect to the equivalence group, the values of arbitrary elements for each of that
the equation~$\mathcal{L}_\theta$ admits additional reduction operators.

In an analogues way we also can introduce equivalence relations on~$P$, which are
generated by either generalizations of usual equivalence groups or
all admissible point transformations~\cite{Popovych&Eshraghi2005MOGRAN}
(called also form-preserving ones~\cite{Kingston&Sophocleous1998})
in pairs of equations from~$\mathcal{L}|_{\cal S}$.

\section{Reduction operators\\ of nonlinear filtration equation}
\label{SectionOnRedOpsOfPotFastDifEq}

In this section we describe $G_2$-inequivalent reduction operators of the potential fast diffusion
equation~\eqref{PotFastDifEq}.
Here reduction operators have the general form
$Q=\tau\p_t+\xi\p_x+\theta\p_v$, where $\tau$, $\xi$ and~$\theta$ are functions of $t$,
$x$ and $v$, and $(\tau,\xi)\not =(0,0)$.
Since \eqref{PotFastDifEq} is an evolution equation, there are two principally different
cases of finding $Q$: $\tau\ne0$ and $\tau=0$.

In the case $\tau=0$ we have $\xi\not=0$, and up to the usual equivalence of reduction operators we can assume
that $\xi=1$, i.e. $Q=\p_x+\theta\p_v$. The conditional invariance criterion implies only one determining equation on
the coefficient~$\theta$
\[
\theta\theta_t=\theta_{xx}+2\theta\theta_{xv}+\theta^2\theta_{vv}
-\theta^{-1}(\theta_x)^2-2\theta_x\theta_v-\theta(\theta_v)^2
\]
which is reduced with a non-point transformation to equation~\eqref{PotFastDifEq},
where $\theta$ becomes a parameter. That is why the case $\tau=0$ is called the ``no-go'' one.
It is characteristic for evolution equations in general.
First the ``no-go'' case was completely investigated for the one-dimensional linear heat equation
in~\cite{Fushchych&Shtelen&Serov&Popovych1992}.
It was proved that the problem of finding the conditional symmetry operators with the vanishing coefficient
of $\p_t$ is reduced to solving the initial equation.
In~\cite{Zhdanov&Lahno1998} the proof was extended to the class of $(1+1)$-dimensional evolution equations
and in~\cite{Popovych1998} this result was generalized for evolution equations with $n$ space variables.

Let us note that ``no-go'' has to be treated as impossibility of exhaustive solving of the problem.
At the same time, imposing additional constraints on the coefficient~$\theta$,
one can construct a number of particular examples of operators with $\tau=0$
and then apply them to finding exact solutions of the initial equation.
It is the approach that was used in~\cite{Gandarias2001} for fast diffusion equation~\eqref{FastDifEq}.
Since the determining equation has more independent variables and, therefore, more freedom degrees,
it is more convenient often to guess a simple solution or a simple ansatz
for the determining equation, which can give a parametric set of complicated solutions of the initial equation.
(Similar situation is for Lie symmetries of first-order ordinary differential equations.)

Consider the case $\tau\ne0$ which admits complete solving unlike the previous case.
We can assume $\tau=1$ up to the usual equivalence of reduction operators.
Then the determining equations for the coefficients $\xi$ and $\theta$ have the form
\begin{gather}\label{EqDetForRedOpsOfPotFastDifEq}\arraycolsep=0ex
\begin{array}{l}
\xi_{vv}=\xi\xi_v, \qquad
\xi_t=2\xi_{xv}-\theta_{vv}-\theta_v\xi+\theta\xi_v-\xi\xi_x,\\[1.5ex]
\theta_{xx}=\theta\theta_x, \qquad
\theta_t=2\theta_{xv}-\xi_{xx}-\xi_x\theta+\xi\theta_x-\theta\theta_v.
\end{array}
\end{gather}

\begin{theorem}\label{TheoremOnRedOpsOfPotFastDifEq}
A complete list of $G_2$-inequivalent non-Lie reduction operators of
the potential fast diffusion equation~\eqref{PotFastDifEq} is exhausted by the following ones:

\bigskip

$1.\ \p_t+\varepsilon\p_x+f(\omega)\p_v,\quad\mbox{where}\quad \omega=x+\varepsilon t;$

\bigskip

$2.\ \p_t+f(\omega)(\p_x+\p_v),\quad\mbox{where}\quad \omega=x+v;$

\bigskip

$3.\ \p_t+\xi\p_x+(\varphi_t+\varphi_x\xi)\p_v,\quad\mbox{where}\quad
\xi=\dfrac{-2}{v+\varphi}, \quad \varphi\in\{t+ e^x,\,t f(x)\};$

\bigskip

$4.\ \p_t+\xi\p_x-\dfrac{\chi_t+\chi_x \xi}{1+\chi^2}\p_v,\quad\mbox{where}\quad \xi=-2\dfrac{1+\chi \tan v}{\tan v-\chi},\ $

\medskip

${}\quad\chi\in\{\tan(2t)\tanh x,\,\coth(2t)\cot x\};$

\bigskip

$5.\ \p_t+\xi\p_x-\dfrac{\chi_t+\chi_x \xi}{1-\chi^2}\p_v,\quad\mbox{where}\quad \xi=-2\dfrac{1-\chi \tanh v}{\tanh v-\chi},$

\medskip

${}\quad\chi\in\Bigl\{\tanh(2t)\tanh x,\,\tanh(2t)\coth x,\,\coth(2t)\coth x,\,
\dfrac{e^{2x}\tanh2t+1}{e^{2x}-\tanh2t},\,\dfrac{2-e^{2x}-e^{4t}}{2+e^{2x}+e^{4t}}\Bigr\}.$

\bigskip

\noindent
Here $\varepsilon\in\{0,\,1\}$,
$f$ is an arbitrary nonconstant solution of the ordinary differential equation \mbox{$f_{\omega\omega}=ff_\omega$}, i.e.
$f\in\{-2/\omega,\,-2\cot\omega,\,-2\tanh\omega,\,-2\coth\omega\}\!\!\!\mod G_2.$
\end{theorem}

\begin{proof}\looseness=-1
Here we only outline a sketch of proof.
Any solution of the equation~$\xi_{vv}=\xi\xi_v$ belongs to the set
$
\{\varphi,\, -2/(v+\varphi),\,-2\mu\cot\omega,\,-2\mu\tanh\omega,\,-2\mu\coth\omega\},
$
where $\omega=\mu(v+\varphi)$, $\mu$~and $\varphi$ are arbitrary functions of~$t$ and~$x$, $\mu\not=0$.
The second equation of~\eqref{EqDetForRedOpsOfPotFastDifEq} is a linear inhomogeneous second-order
ordinary differential equation with respect to~$\theta$, where $v$ is the independent variable and
$t$ and $x$ are assumed parameters.
It is possible to construct its partial exact solution without irrational singularities for
any above value of $\xi$.
The solutions of the corresponding homogeneous equation have irrational singularities if $\xi_v\not=0$.
In view of the other equations of~\eqref{EqDetForRedOpsOfPotFastDifEq},
the part of~$\theta$ containing such singularities has to vanish identically.
Moreover, $\mu=\const$, i.e. $\mu=1\!\!\mod G_2$, and $\varphi$ satisfies an overdetermined system
of differential equation in~$t$ and~$x$.
Integration of it for all above values of $\xi$ and classification of obtained solutions
up to equivalence with respect to $G_2$ with excluding Lie cases result in the statement of the theorem.
\end{proof}

All operators from Theorem~1 are potential nonclassical symmetries of equation~\eqref{FastDifEq}.


\section{Connection between classes of\\ nonclassical and potential nonclassical symmetries}
\label{SectionOnConnectionBetweenUsualAndPotRedOpsOfFastDifEq}

Let us investigate connection between reduction operators of equations~\eqref{FastDifEq} and~\eqref{PotFastDifEq}.

As mentioned in the introduction, one of the problems studied in~\cite{Gandarias2001} was
construction of partial classes of reduction operators for diffusion equations~\eqref{diffusionmodel}
in the ``no-go'' case when operators can be reduced to the form $\p_x+\eta(t,x,u)\p_u$.
Namely, Gandarias proposed to look for the coefficient~$\eta$ with the ansatz
\begin{equation}\label{EqAnsatzForEtaInNoGoCaseNDEs}
\eta=\dfrac{\eta^1(t,x)u+\eta^2(t,x)}{f(u)},
\end{equation}
where  $f(u)\equiv u^{-\alpha}$.
After substituting~\eqref{EqAnsatzForEtaInNoGoCaseNDEs} in the determining equation for~$\eta$ and
splitting with respect to~$u$, one obtains an overdetermined system for the functions~$\eta^1$ and~$\eta^2$.
For the fast diffusion equation~\eqref{FastDifEq} this system has the form
\begin{equation}\label{EqForEta1Eta2InNoGoCaseNDEs}
\eta^2_{xx}=\eta^2\eta^2_x,\qquad
\eta^2_t=\eta^2\eta^1_x-\eta^1\eta^2_x+\eta^1_{xx},\qquad
\eta^1_t=\eta^1\eta^1_x.
\end{equation}
System~\eqref{EqForEta1Eta2InNoGoCaseNDEs} can be derived from~\eqref{EqDetForRedOpsOfPotFastDifEq} with reduction
by the group of translations with respect to~$v$, i.e. with assuming
that $\xi$ and~$\theta$ do not depend on~$v$ and re-denoting $\xi=-\eta^1$, $\theta=\eta^2$.

This observation can be easily explained in a rigorous way for any pair of
equations~\eqref{EqNonlinDiff} and~\eqref{EqNonlinFiltration} with the same function~$f$.

Consider reduction operators $Q=\p_t+\xi\p_x+\theta\p_v$ and $Q'=\p_x+\eta\p_u$
of equations~\eqref{EqNonlinFiltration} and~\eqref{EqNonlinDiff} correspondingly,
where the coefficients~$\xi$ and~$\theta$ depend only on~$t$ and~$x$,
the coefficient~$\eta$ is defined by~\eqref{EqAnsatzForEtaInNoGoCaseNDEs}.
The conditional invariance criterion applied to equation~\eqref{EqNonlinFiltration} (or \eqref{EqNonlinDiff}\,)
and the operator~$Q$ ($Q'$) implies the following determining equation on~$\xi$ and~$\theta$ ($\eta^1$ and~$\eta^2$):
\begin{gather*}
(\xi\xi_xv_x{}^2-(\xi_x\theta+\xi\theta_x)v_x+\theta\theta_x)f'(v_x)
\\
\qquad+((\xi_t+2\xi\xi_x)v_x-\theta_t-2\theta\xi_x)f(v_x)+(-\xi_{xx}v_x+\theta_{xx})(f(v_x))^2=0
\\[1ex]
(\mbox{or}\quad
(\eta^1u+\eta^2)(\eta^1_xu+\eta^2_x)f'(u)-
\\
\qquad((\eta^1_t-2\eta^1\eta^1_x)u+\eta^0_t-2\eta^0\eta^1_x)f(u)+(\eta^1_{xx}u+\eta^2_{xx})(f(u))^2=0
\ )
\end{gather*}
which has to be additionally split with respect to $v_x$ ($u$).
It is obvious that the systems obtained in the both cases after splitting coincide under
the supposition $\eta^1=-\xi$, $\eta^2=\theta$.
The characteristic equation $Q[v]=\theta-v_t-\xi v_x=0$ can be rewritten
on the manifold of solutions of the potential system
\begin{equation}\label{EgPotSystemForNDEs}
v_x=u,\quad v_t=f(u)u_x
\end{equation}
in the form
\[u_x-\frac{-\xi u+\theta}{f(u)}=0\]
and coincides in this way with the characteristic equation~$Q'[u]=0$.

Therefore, the following proposition is true.

\begin{proposition}
$Q=\p_t+\xi\p_x+\theta\p_v$, where $\xi=\xi(t,x)$ and $\theta=\theta(t,x)$,
is a reduction operator of equation~\eqref{EqNonlinFiltration} iff
\[Q'=\p_x+\frac{-\xi u+\theta}{f(u)}\p_u\]
is a reduction operator of equation~\eqref{EqNonlinDiff}.

System~\eqref{EgPotSystemForNDEs} establishes connection between the corresponding sets of invariant solutions.
\end{proposition}

\section{Exact solutions of fast diffusion\\ and nonlinear filtration equations}
\label{SectionOnSolutionOfFastDifEq}

All invariant solutions of~\eqref{FastDifEq}
and~\eqref{PotFastDifEq}, which were earlier constructed in closed
forms with the classical Lie method, were collected e.g.
in~\cite{Polyanin&Zaitsev2004,Popovych&Ivanova2005PETs}. A
complete list of $G_1$-inequivalent solutions of such type is
exhausted by the following ones:
\begin{gather}\label{Liesolutions.for.u-1}
\begin{split}&
1)\ u=\dfrac{1}{1+\varepsilon e^{x+t}},\quad  v=-\ln|e^{-x}+\varepsilon e^t|;\\[1ex]&
2)\ u=e^{x},\quad v=e^x+t;\\&
3)\ u=\dfrac{1}{x-t+\mu te^{-x/t}},\quad
v=\ln|t|+\left.\int\!\!\frac{d\vartheta}{\vartheta-1+\mu e^{-\vartheta}}\,\right|_{\vartheta=x/t};
\\[1ex]&
4)\ u=\dfrac{2t}{x^2+\varepsilon t^2},\quad v\bigr|_{\varepsilon=0}=-\frac{2t}{x},\quad
v\bigr|_{\varepsilon=1}=2\arctan \frac xt,\quad v\bigr|_{\varepsilon=-1}=\ln\left|\frac{x-t}{x+t}\right|;\\[1ex]&
5)\ u=\dfrac{2t}{\cos^2x},\quad v=2t\tan x;\\[1ex]&
6)\ u=\dfrac{-2t}{\cosh^2x},\quad v=-2t\tanh x;\\[1ex]&
7)\ u=\dfrac{2t}{\sinh^2x},\quad v=-2t\coth x.
\end{split}\end{gather}
Here $\varepsilon$ and $\mu$ are arbitrary constants, $\varepsilon\in\{-1,0,1\}\!\!\mod G_1\!$.
The below arrows denote the possible transformations of solutions~\eqref{Liesolutions.for.u-1}
to each other by means of the potential hodograph transformation~\eqref{pothodograph}
up to translations with respect to~$x$~\cite{Popovych&Ivanova2005PETs}:
\[
\begin{split}&
\mbox{\Large$\circlearrowright$}\;1)_{\varepsilon=0}\,; \quad
1)_{\varepsilon=1}\longleftrightarrow 1)_{\varepsilon=-1,\;x+t<0}\,; \quad
\mbox{\Large$\circlearrowright$}\;1)_{\varepsilon=-1,\;x+t>0}\,; \quad
2)\longleftrightarrow 3)_{\mu=0,\;x>t}\,; \\&
\mbox{\Large$\circlearrowright$}\;4)_{\varepsilon=0}\,; \quad
5)\longleftrightarrow 4)_{\varepsilon=4}\,; \quad
6)\longleftrightarrow 4)_{\varepsilon=-4,\;|x|<2|t|}\,; \quad
7)\longleftrightarrow 4)_{\varepsilon=-4,\;|x|>2|t|}\,.
\end{split}
\]
The sixth connection was known earlier~\cite{Fushchych&Serov&Amerov1982,Pukhnachov1996}.
If $\mu\ne0$ solution 3) from list~\eqref{Liesolutions.for.u-1} is mapped by~\eqref{pothodograph}
to the solution
\[
8)\ u=t\vartheta(\omega)-t+\mu te^{-\vartheta(\omega)},\qquad \omega=x-\ln|t|,
\]
which is invariant with respect to the algebra~$\langle t\p_t+\p_x+u\p_u\rangle$.
Here $\vartheta$ is the function determined implicitly by the formula
$
\int (\vartheta-1+\mu e^{-\vartheta})^{-1}d\vartheta=\omega.
$

Some classes of non-Lie exact solutions of~\eqref{FastDifEq} were obtained in~\cite{Rosenau1995,Qu1999,Gandarias2001}.
These solutions and the ones similar to them can be represented uniformly over the complex field
as compositions of two simple waves moving with the same ``velocities'' in opposite directions:
\begin{gather}\label{TanRepresentation.for.u-1}
\begin{split}&
u=\frac{\alpha^2}{\beta}(-\cot(\alpha x+\beta t+\gamma)+\cot(\alpha x-\beta t+\delta))
\\[1ex]&
\phantom{u}=\frac{\alpha^2}{\beta}
\frac{2\sin(2\beta t+\gamma-\delta)}{\cos(2\beta t+\gamma-\delta)-\cos(2\alpha x+\gamma+\delta)},
\end{split}
\end{gather}
where $\alpha$, $\beta$, $\gamma$ and $\delta$ are complex constants, $\alpha\beta\not=0$.
It can be proved that function~\eqref{TanRepresentation.for.u-1} takes real values (for real $x$ and $t$) iff
up to transformations from~$G_1$
\[
(\alpha,\beta,\gamma,\delta)\in
\{(1,1,0,0),\, (i,i,0,0),\, (i,i,\pi/2,0),\, (i,i,\pi/2,\pi/2),\, (i,1,0,0),\, (1,i,0,0)\}.
\]
Using representation~\eqref{TanRepresentation.for.u-1} and the above values of tuples $(\alpha,\beta,\gamma,\delta)$,
we obtain the following solutions of fast diffusion equation~\eqref{FastDifEq} and
nonlinear filtration equation~\eqref{PotFastDifEq}:
\[\begin{split}&
1')\ u=\cot(x-t)-\cot(x+t)=\dfrac{2\sin 2t}{\cos 2t-\cos 2x},\quad v=\ln\left|\frac{\sin(x-t)}{\sin(x+t)}\right|;
\\&
2')\ u=\coth(x-t)-\coth(x+t)=\dfrac{2\sinh 2t}{\cosh2x-\cosh 2t},\quad
v=\ln\left|\frac{\sinh(x-t)}{\sinh(x+t)}\right|;\\[1ex]&
3')\ u=\coth(x-t)-\tanh(x+t)=\dfrac{2\cosh 2t}{\sinh 2x-\sinh 2t},
\quad v=\ln\left|\frac{\sinh(x-t)}{\cosh(x+t)}\right|;\\[1ex]&
4')\ u=\tanh(x-t)-\tanh(x+t)=-\dfrac{2\sinh 2t}{\cosh 2x+\cosh 2t},\quad
v=\ln\left|\frac{\cosh(x-t)}{\cosh(x+t)}\right|;\\[1ex]&
5')\ u=\cot(ix+t)-\cot(ix-t)=\dfrac{2\sin 2t}{\cosh 2x-\cos 2t},\quad v=2\arctan(\cot t\,\tanh x);\\[1ex]&
6')\ u=i\cot(x+it)-i\cot(x-it)=\dfrac{2\sinh 2t}{\cosh 2t-\cos 2x},\quad v=2\arctan(\coth t\,\tan x).
\end{split}\]
(All two's in the latter expressions for $u$ can be moved over with scale transformations from~$G_1$.)

Transformation~(\ref{pothodograph}) acts on the set of solutions $1')$--$6')$
in the following way:
\[
\begin{split}&
1')_{\cos 2t<\cos 2x} \longleftrightarrow 5')|_{t\to t+\pi/2,\,x\to x/2,\,v\to 2v}; \quad
1')_{\cos 2t>\cos 2x} \longleftrightarrow 5')|_{x\to x/2,\,v\to 2v-\pi}\,;
 \quad
\\&
2')_{|x|<|t|}\longleftrightarrow 4')|_{x\to x/2,\,v\to 2v}; \quad
\mbox{\Large$\circlearrowright$}\;2')_{|x|>|t|}|_{x\to x/2,\,v\to 2v}\,;\quad
\\&
\mbox{\Large$\circlearrowright$}\;3')_{x<t}|_{x\to x/2,\,v\to 2v}; \quad
3')_{x>t} \longleftrightarrow 3')_{x>t}|_{x\to -x/2,\,v\to -2v}; \quad
\mbox{\Large$\circlearrowright$}\;6')|_{x\to x/2,\,v\to 2v}
.
\end{split}
\]
These actions can be interpreted in terms of actions of transformation~(\ref{pothodograph}) on
the nonclassical symmetry operators which correspond to solutions $1')$--$6')$.

In~\cite{Rosenau1995} Rosenau took advantage of additive separation of variables
for the potential fast diffusion equation~\eqref{PotFastDifEq} and constructed solution $4')$.
Using the generalized conditional symmetry method, Qu~\cite{Qu1999} found
solutions which can be written in forms $1')$ and $6')$.
After rectifying computations in two cases from~\cite{Qu1999}, one
can find also solutions $2')$ and $5')$.
Solutions $1')$, $3')$ and $4')$ were obtained in \cite{Gandarias2001}, at least, in one from the above forms,
but equivalence of these forms was not shown there.

One of techniques which can be applied for finding the above solutions is reduction by conditional symmetry operators
of the form $Q=\p_x+(\eta^1(t,x)u+\eta^2(t,x))u\p_u$ (see~\cite{Gandarias2001} for details).
Namely,  solutions $1')$, $3')$ and $4')$ are obtained with the following operators:
\[
\p_x+(u^2-2\cot(x-t)u)\p_u,\quad\p_x+(u^2-2\coth(x-t)u)\p_u\quad\mbox{and}\quad\p_x+(u^2-2\tanh(x-t)u)\p_u.
\]
We supplement the list of solutions adduced in~\cite{Gandarias2001,Qu1999,Rosenau1995} with similar ones,
namely, with $2')$ and $5')$.
Solution $2')$ can be also constructed by reduction with the second above operator.
Real solutions $5')$ and $6')$ correspond to the similar operators
\begin{gather*}
\p_x+(iu^2-2\coth(x-it)u)\p_u \quad\mbox {and} \quad \p_x-(iu^2-2i\coth(t-ix)u)\p_u
\end{gather*}
with complex-valued coefficients.

All reductions performed with reduction operators from Theorem 1 or with equivalent ones
result in solutions which are equivalent to the listed Lie solutions 1)--7) or solutions $1')$--$6')$.
For example, the operators
\[
\p_t-\p_x-2\cot(x-t)\p_v, \quad \p_t-\p_x-2\coth(x-t)\p_v \quad\mbox {and} \quad \p_t-\p_x-2\tanh(x-t)\p_v
\]
lead to solutions $1')$, $3')$ and $4')$ correspondingly
(see also Section~\ref{SectionOnConnectionBetweenUsualAndPotRedOpsOfFastDifEq}).

\section{Conclusion}

In this Letter we present classification of reduction operators for
nonlinear filtration equation~\eqref{PotFastDifEq} with summary of necessary notions and statements,
a basic sketch of the proof and a list of constructed exact solutions including both Lie and non-Lie ones.
Since \eqref{PotFastDifEq} is the potential equation of fast diffusion equation~\eqref{FastDifEq},
all the obtained operators are potential nonclassical symmetries of~\eqref{FastDifEq}.
Moreover, most of them are nonprojectible on the space of the independent variables~$t$ and $x$ that leads to
technically cumbersome implicit reductions of \eqref{PotFastDifEq} to ordinary differential equations.
Now we optimize the proof and hope to realize it in a quite compact form.
Presentation of the proof and reduction technics will be subjects of a future paper.

We continue our investigation on potential reduction operators of the nonlinear diffusion equations
from class~\eqref{EqNonlinDiff}.
In some sense, equation~\eqref{FastDifEq} is singular in this class with the potential nonclassical symmetry
point of view.
More precisely, as a result of joint work with Prof.~Sophocleous the following theorem have been proved recently.

\begin{theorem}\label{TheoremOnPotRedOpsOfDifEq}
Nonlinear filtration equations~\eqref{EqNonlinFiltration}
admit non-Lie reduction operators with non-vanishing coefficients of~$\p_t$
only in the case of the Fujita's nonlinearities
\[
f(v_x)=\dfrac1{av_x{}^2+bv_x+c},
\]
where $a$, $b$ and $c$ are constants.
\end{theorem}

Let us note that there are exactly three $G^{\Equiv}$-inequivalent cases of the Fujita's nonlinearities:
\[
f(v_x)=1,\quad f(v_x)=\dfrac1{v_x},\quad f(v_x)=\dfrac1{v_x^2+1}.
\]
The equivalence group $G^{\Equiv}$ of class~\eqref{EqNonlinFiltration} is formed by the transformations
\begin{gather*}
\tilde t=\varepsilon_1t+\varepsilon_2,\quad
\tilde x=\varepsilon_1'x+\varepsilon_2'v+\varepsilon_3',\quad
\tilde v=\varepsilon_1''x+\varepsilon_2''v+\varepsilon_3'',\quad
\tilde f=\varepsilon_1^{-1}(\varepsilon_1'+\varepsilon_2'v_x)^2\,f,
\end{gather*}
where $\varepsilon_1,$ $\varepsilon_2,$ $\varepsilon_i',$ $\varepsilon_i''$ $(i=1,2,3)$ are arbitrary constants,
$\varepsilon_1(\varepsilon_1'\varepsilon_2''-\varepsilon_2'\varepsilon_1'')\ne0.$
The nonclassical (conditional) symmetries of the $(1+1)$-dimensional linear heat equation ($f=1$) were completely studied
in~\cite{Fushchych&Shtelen&Serov&Popovych1992}.
Analogous investigation of the second case ($f=v_x^{-1}$) is carried out in this Letter. 
Therefore, to complete  classification of reduction operators in the class of
nonlinear filtration equations~\eqref{EqNonlinFiltration} with respect to~$G^{\Equiv}$
(see Definition~\ref{DefinitionOfEquivOfRedOperatorsWrtEquivGroup}),
it is enough to describe reduction operators
of the equation with the latter nonlinearity, and we achieved significant progress in solving this problem.

\subsection*{Acknowledgements}

The authors are grateful to Prof. C.~Sophocleous for useful discussions and interesting comments.
ROP and OOV thank University of Cyprus for hospitality and support during writing this paper.
The research of ROP was supported by Austrian Science Fund (FWF), Lise Meitner
project M923-N13.
The research of OOV and NMI was partially supported by the grant of the President of Ukraine for
young scientists GF/F11/0061.
NMI acknowledges financial support from National Sciences and Engineering Council of Canada and
Department of Mathematics of the University of British Columbia.
The authors also wish to thank the referees for careful
reading and suggestions for improvement of this Letter.

\end{document}